\documentclass[aps,12pt]{revtex4}

\usepackage{amssymb, amsmath, hyperref}
\usepackage[final]{graphicx}
\usepackage[latin1]{inputenc}

\begin{document}

\rightline{SPIN-06-15, ITP-UU-06-18}

\title{Instabilities in the nonsymmetric theory of gravitation}

\author{Tomas Janssen\footnote{T.Janssen2@phys.uu.nl}
        and Tomislav Prokopec\footnote{T.Prokopec@phys.uu.nl}}

\affiliation{Institute for Theoretical Physics, University of Utrecht
             Leuvenlaan 4, Postbus 80.195,
              3508 TD Utrecht, The Netherlands}

\begin{abstract}
We consider the linearized nonsymmetric theory of gravitation
(NGT) within the background of an expanding universe and near a
Schwarzschild metric. We show that the theory always develops
instabilities unless the linearized nonsymmetric lagrangian
reduces to a particular simple form. This theory contains a gauge
invariant kinetic term, a mass term for the antisymmetric
metric-field and a coupling with the Ricci curvature scalar. This
form cannot be obtained within NGT. Next we discuss NGT beyond
linearized level and conjecture that the instabilities are not a
relic of the linearization, but are a general feature of the full
theory. Finally we show that one cannot add {\it ad-hoc}
constraints to remove the instabilities as is possible with the
instabilities found in NGT by Clayton.

\end{abstract}

\maketitle

\section{Introduction}

So far, Einstein's general relativity  (GR) has stood all direct
experimental tests. The precession of the perihelium of Mercury,
gravitational lensing and the redshift of light are measured in
agreement with the theory at the percent level, just to mention a
few. The crown on this experimental evidence, a direct measurement
of gravitational waves, is expected within a few years
~\cite{Will:2005va}.

 However there are also reasons to try to extend GR.  For example the mysterious nature of dark energy and dark matter might become
resolved within a modified theory of gravity.
The standard example of a modified theory of gravity
 is Milgrom's Modified Newtonian Dynamics
(MOND)~\cite{Milgrom:1983ca}, according to which Newton's law gets
modified at very low accelerations, presenting an alternative to
dark matter. MOND has recently been covariantized by
Bekenstein~\cite{Bekenstein:2004ne}.

 Another reason to try to extend GR is the notion of generality.
Within the framework of GR torsion is not included in a natural,
geometric way. Within standard GR, any calculation of the
connection (either by requiring metric compatibility, or by using
the first order formalism) leads to the (symmetric) Levi-Civit\`a
connection. Of course one is then free to add torsion, but there
is no way that torsion follows naturally from the theory. An
interesting generalization of GR would generate torsion in a
purely geometric way, analogous to the way the Levi-Civit\`a
connection is generated in GR.

 The Nonsymmetric Gravitational Theory (NGT)~\cite{Moffat:1994hv}
is an extension of GR that drops the standard axiom of GR that the
metric is a symmetric tensor. Therefore we write the general,
nonsymmetric metric as
\begin{equation} \label{naive}
g_{\mu\nu}=G_{\mu\nu}+B_{\mu\nu},
\end{equation}
where $G_{\mu\nu}=g_{(\mu\nu)}$, $B_{\mu\nu}=g_{[\mu\nu]}$ and
$(\cdot)$ and $[\cdot]$ indicate normalized symmetrization and
anti-symmetrization, respectively, as defined in
Eq.~(\ref{definitions}). Indeed there is no physical principle
that tells us that the metric should be symmetric and therefore
such a generalization is very interesting to study. These kinds of
theories where originally studied by
Einstein~\cite{Einstein:1945}~\cite{Einstein:1974} in an attempt
to unify GR with Maxwell's theory. However it turned out to be
impossible to regain the Lorentz force
law~\cite{Infeld:1950}~\cite{callaway:1953}. The theory was
revived by Moffat~\cite{Moffat:1978tr}, but now with the
interpretation that the antisymmetric part of the metric really
produces a new gravity-like force. This extra structure produces
interesting results on the issues of dark energy and dark
matter~\cite{Moffat:2004bm}~\cite{Moffat:2004nw}~\cite{ProkopecWessel:2006}.
It will be clear that such a theory produces torsion in a very
natural way. Furthermore note that a dynamical nonsymmetric metric
field also naturally arises from non-linear $\sigma$-models of
string theory~\cite{Kalb:1974yc}. Unfortunately the nonsymmetric
theory of gravitation suffers from all kinds of problems. The
first of these is the notion of non-uniqueness. Because of the
extra structure in NGT, there is no natural choice for the
lagrangian. Without a guiding principle, the most general
lagrangian that one can write down (with up to two time
derivatives) has 9 extra terms apart from the usual Ricci scalar
term (see Eq.~(\ref{lag})). Non-uniqueness also comes in another
way: if we consider the expansion of the theory in terms of the
$B$-field, we could in general write
\begin{equation}
 \label{decomposition}
\begin{split}
    g_{\mu\nu}&=G_{\mu\nu}+B_{\mu\nu}
               +\rho B_{\mu\alpha}B^\alpha{}_\nu+\sigma B^2 G_{\mu\nu}
               +\mathcal{O}(B^3)\\
    g^{\mu\nu}&=G^{\mu\nu}+B^{\mu\nu}
               +(1-\rho) B^{\mu\alpha}B_\alpha{}^\nu+\sigma B^2 G^{\mu\nu}
               +\mathcal{O}(B^3),
\end{split}
\end{equation}
where $B^2=G^{\mu\alpha}G^{\nu\beta}B_{\mu\nu}B_{\alpha\beta}$ and
$B_\alpha{}^\nu=G^{\rho\nu}B_{\alpha\rho}$. $\rho$ and $\sigma$
are undetermined parameters. The inverse is defined by
$g^{\mu\alpha}g_{\mu\beta}=\delta^\alpha_\beta$. With these two
extra parameters one sees that the linearized lagrangian is
determined by 11 undetermined parameters coming from the full
theory and the decomposition of the metric
tensor~(\ref{decomposition}).

Another problem with NGT, found by Damour, Deser and
McCarthy~\cite{Damour:1992bt}, is the absence of gauge invariance
\begin{equation} \label{gauge}
    B_{\mu\nu}\rightarrow B_{\mu\nu}+\partial_\mu\lambda_\nu-\partial_\nu\lambda_\mu
\end{equation}
for the $B$-field. This absence occurs, since in the linearized
expansion one obtains, apart from the gauge invariant kinetic term
\begin{equation}
-\frac{1}{12}H^2\quad\quad;\quad\quad
H_{\alpha\beta\gamma}=\partial_\alpha
B_{\beta\gamma}+\partial_\beta B_{\gamma\alpha}+\partial_\gamma B_{\alpha\beta}
\end{equation}
also undifferentiated powers of $B$ that couple to the background
curvature. The absence of gauge invariance might lead to the
propagation of ghosts, or unacceptable constraints on dynamical
degrees of freedom. Fortunately these problems can be relatively
easily solved by the introduction of a mass term for the
$B$-field~\cite{Moffat:1994hv}~\cite{Clayton:1995yi}.

The final problem we discuss was found by
Clayton~\cite{Clayton:1996dz}. One can show that, if one starts
with a GR configuration and a small $B$-field, the $B$-field will
quickly grow and therefore the linearization of the $B$-field does
not make much sense. Since the field equations for the symmetric
and antisymmetric part of the metric decouple at linearized level,
Clayton argues that the instabilities might be a property of the
linearized theory only. At higher orders, the $B$-field backreacts
on the symmetric metric and therefore the GR background should be
considered as evolving. Such an evolving background leads to an
increase of the degrees of freedom of the
$B$-field~\cite{Isenberg:1977} and these extra degrees of freedom
then might stabilize the field. Another (phenomenological)
solution to these instabilities is given by
Moffat~\cite{Moffat:1996hf}. The idea of this solution is to
introduce an extra Lagrange multiplier in order that the unstable
modes dynamically vanish.

 For the moment forgetting Clayton's result, in this paper we
consider NGT linearized around a GR configuration. By explicitly
constructing two different backgrounds (FLRW-universe in
section~\ref{scosmo}, and Schwarzschild metric in
section~\ref{sschwarz}) we show that the evolution of the
$B$-field is indeed unstable. By considering the most general form
of the linearized lagrangian, we can explicitly point out which
terms cause these instabilities. After having shown this we argue
in section \ref{ssextend} that instabilities are also present
beyond linear level. Therefore our results can be extended to the
full theory and should not be seen as a relic of the linearized
theory. If these arguments are correct we must then conclude that
even full, nonlinearized NGT suffers from instabilities. Finally
we show in section \ref{sdynamically} that the dynamical solution
of Moffat to solve the instabilities found by Clayton does not
solve the instabilities we find.

 In section~\ref{slin} we briefly summarize the linearized theory,
in sections~\ref{scosmo} and~\ref{sschwarz} we evaluate the
evolution of the $B$-field in two different GR backgrounds. In
section~\ref{sdisc} we analyze the results, discuss the theory
beyond linear level and consider dynamically constrained NGT. In
section~\ref{scon} we draw our conclusions. In appendix \ref{alin}
we state our notation and give the derivation of the linearized
lagrangian. Appendix~\ref{ageom} summarizes the geometric
quantities needed in the text.

\section{The linearized Lagrangian} \label{slin}
Since GR is very successful, it is natural to assume that any
modification of the theory should be relatively small. Therefore
we consider NGT in the limit of a small $B$, but an arbitrary $G$.
The linearized Lagrangian of such a theory will in general have
the following form (see Appendix~\ref{alin}):
\begin{equation} \label{lagrangian}
\begin{split}
    \mathcal{L}=\sqrt{-G}\bigg[&R+2\Lambda-\frac{1}{12}H^2+(\frac{1}{4}m^2+\beta R)B^2\\&-\alpha R_{\mu\nu}B^{\mu\alpha}B_\alpha{}^\nu-\gamma R_{\mu\alpha\nu\beta}B^{\mu\nu}B^{\alpha\beta}\bigg]+\mathcal{O}(B^3).
\end{split}
\end{equation}
Here the curvature terms $R_{\mu\alpha\nu\beta}$, $R_{\mu\nu}$ and
$R$ all refer to the background, GR, curvature. The coefficient
$\gamma$ is determined by the particular choice of the 'full'
lagrangian, $\beta$ is determined by the decomposition of the
metric (\ref{decomposition}), while $\alpha$ depends both on the
full lagrangian and the decomposition. Naturally, different
choices of these coefficients lead to different physical theories,
so though in principle we can always make a decomposition of the
metric in such a way that $\alpha$ and $\beta$ vanish, we have no
guiding principle that tells us they \emph{must} be zero and
therefore we will keep them arbitrary. The coefficient $\gamma$
cannot be set to zero within the first order (Palatini) formalism
(see appendix \ref{alin}). Since the curvature couplings break the
gauge invariance (\ref{gauge}), the mass $m^2$ has been added to
prevent ghost modes to propagate ~\cite{Damour:1992bt}. This is
not as artificial as it may sound. A mass is
natural, since it is automatically generated in the presence of a
nonzero cosmological constant (see Eq. (\ref{mass})). We may assume
that the mass term is generated by a cosmological constant
and therefore we have today,
\begin{equation}
    m^2\sim \Lambda\leq 10^{-84}~{\rm GeV}^2
.
\label{assumption}
\end{equation}
Note that this inequality is not necessarily true at all times,
especially not when one considers an epoch of the early Universe,
since the cosmological term may change during the evolution of the
Universe (for example during phase transitions). The field
equations derived from the lagrangian (\ref{lagrangian}) are
\begin{equation} \label{fieldeq}
\begin{split}
&(\sqrt{-G})^{-1}\frac{1}{2}\partial_\rho (\sqrt{-G}H^{\rho\mu\nu})
 + (\frac{1}{2}m^2+2\beta R)B^{\mu\nu}\\
&\qquad\qquad -\alpha(B^{\nu\alpha}R^\mu{}_\alpha
  +B^{\alpha\mu}R^\nu{}_\alpha)-2\gamma
B^{\alpha\beta}R^\mu{}_\alpha{}^\nu{}_\beta +\mathcal{O}(B^2)=0\\
&R_{\mu\nu}-\frac12 R G_{\mu\nu}-\Lambda G_{\mu\nu}+\mathcal{O}(B^2)=0
\,.
\end{split}
\end{equation}
We therefore see that to this order the field equations decouple
and it makes sense to consider the symmetric background, to be
just a GR background. The problem then reduces to the study of an
antisymmetric tensor field coupled to GR.

\section{The cosmological evolution of the B-field}
\label{scosmo}

In this section we study the behavior of the B-field in an
expanding universe ~\cite{Prokopec:2005fb}. Our background metric
is given by the usual (conformal)
Friedmann-Lemaitre-Robertson-Walker metric (FLRW):
\begin{equation} \label{metric}
    G_{\mu\nu}=a(\eta)^2\eta_{\mu\nu},
\end{equation}
where $\eta_{\mu\nu}=\text{diag}(1,-1,-1,-1)$, $\eta$ is conformal
time and $a(\eta)$ is the conformal scale factor. The conformal
time is related to the standard cosmological time by, $ad\eta=dt$.
The scale factor during the different cosmological eras is given
in table \ref{tabelletje}, where $H_I \sim 10^{13}~{\rm GeV}$ is the
Hubble parameter during inflation and $\eta_e$ is the conformal
time at matter-radiation equality. For the rest of this section it
is important to note that Eq.~(\ref{assumption}) implies that
\begin{equation}
    H_I\gg m
\,.
\end{equation}
Of course, this bound can be amply satisfied even when the $B$-field mass,
$m$, is much greater than what is indicated in Eq.~(\ref{assumption}).
\begin{table}
\caption{The scale factor and conformal time in different eras}
\label{tabelletje}
\begin{center}
\begin{tabular}{|c|c|c|}
  \hline
  era & $a$ & $\eta$ \\
  \hline
  de Sitter inflation & $a=-\frac{1}{H_I\eta}$ &$\eta\leq-\frac{1}{H_I}$\\
  Radiation & $a=H_I\eta$ & $\frac{1}{H_I}\leq\eta\leq\eta_e$\\
  Matter & $a=\frac{H_I}{4\eta_e^2}(\eta+\eta_e)^2$ &$\eta \geq \eta_e$\\
  \hline
\end{tabular}
\end{center}
\end{table}
\subsection{The lagrangian and the field equations}
The linearized Lagrangian for the $B$-field reads in conformal
metric,
\begin{equation}
\begin{split}
\mathcal{L}=&-\frac{1}{12}\frac{1}{a^2}\eta^{\alpha\beta}\eta^{\rho\sigma}\eta^{\mu\nu}H_{\alpha\rho\mu}H_{\beta\sigma\nu}+(\beta R+\frac{1}{4}m^2)B_{\alpha\beta}B_{\rho\sigma}\eta^{\alpha\rho}\eta^{\beta\sigma}\\
&-\frac{\alpha}{a^2}R_{\mu\nu}B_{\lambda\rho}B_{\alpha\sigma}\eta^{\lambda\mu}\eta^{\rho\alpha}\eta^{\sigma\nu}-\frac{\gamma}{a^4}R_{\mu\alpha\nu\beta}B_{\rho\sigma}B_{\lambda\kappa}\eta^{\rho\mu}\eta^{\sigma\nu}\eta^{\lambda\alpha}\eta^{\kappa\beta}.
\end{split}
\end{equation}
Here the geometric tensors, $R$, $R_{\mu\nu}$, $R_{\mu\nu\rho\sigma}$,
derived from the metric~(\ref{metric})
are given in appendix \ref{ageom} and $\mathcal{H}$
is the conformal Hubble parameter given by
$\mathcal{H}={a'}/{a}$, where $a^\prime = da/d\eta$, denotes a derivative
with respect to conformal time. This lagrangian leads to the
following equations of motion:
\begin{equation} \label{conformaleom}
\begin{split}
&\eta^{\rho\sigma}\partial_\rho H_{\sigma\alpha\beta}-2\mathcal{H}H_{0\alpha\beta}+2a^2\Big(2\beta \mathcal{R}+\frac{1}{2}m^2\Big)B_{\alpha\beta}\\
&-2a^2\alpha(\mathcal{R}^\lambda{}_\alpha
B_{\beta\lambda}+\mathcal{R}^\lambda{}_\beta
B_{\lambda\alpha})-4a^2\gamma
\mathcal{R}^\lambda{}_\alpha{}^\kappa{}_\beta B_{\lambda\kappa}=0
\,.
\end{split}
\end{equation}

Next we act with $\eta^{\alpha\lambda}\partial_\lambda$ on
(\ref{conformaleom}) divided by $2a^2$ to obtain
\begin{equation} \label{consistency}
\begin{split}   &a^{-2}\bigg(-4\alpha\mathcal{H}\mathcal{H}'+(12\beta+4\alpha\mathcal{H}'')-(24\beta+4\alpha)\mathcal{H}^3\bigg)B_{\beta0}\\
&+\bigg(-12\beta(\mathcal{H}^2+\mathcal{H}')+\frac{1}{2}a^2m^2\bigg)a^{-2}\eta^{\alpha\lambda}\partial_\lambda B_{\alpha\beta}\\
&-\alpha\bigg(\mathcal{R}^\lambda{}_\alpha\eta^{\alpha\rho}\partial_\rho
B_{\beta\lambda}+\mathcal{R}^\lambda{}_\beta\eta^{\alpha\rho}\partial_\rho
B_{\lambda\alpha}\bigg)-2\gamma\eta^{\alpha\rho}\partial_\rho\bigg(a^{-2}\eta^{\kappa\sigma}\mathcal{R}^\lambda{}_{\alpha\sigma\beta}B_{\lambda\kappa}\bigg)=0.
\end{split}
\end{equation}
The $0$ and the $i$ components of this equation lead to the
following two consistency conditions ($i=1,2,3$)
\begin{eqnarray}
    &\mathcal{X}\eta^{ij}\partial_j B_{0j}=0 \label{constrain1}\\
    &\partial_0(\mathcal{X}B_{j0})+\mathcal{Y}\eta^{ik}\partial_k B_{ji}=0 \label{constrain2},
\end{eqnarray}
where we defined
\begin{eqnarray}
&\mathcal{X}=a^{-2}\bigg((12\beta+2\alpha)\mathcal{H}^2+(12\beta+4\alpha-2\gamma)\mathcal{H}'-\frac{1}{2}m^2a^2\bigg) \label{X}\\
&\mathcal{Y}=a^{-2}\bigg((12\beta+4\alpha-2\gamma)\mathcal{H}^2+(12\beta+2\alpha)\mathcal{H}'-\frac{1}{2}m^2a^2\bigg)
\label{Y}.
\end{eqnarray}
Next we consider the equations of motion (\ref{conformaleom}). The
$0i$ and the $ij$ components give
\begin{eqnarray}
    &a^{-2}\eta^{\rho\sigma}\partial_\rho\bigg(\partial_\sigma B_{0i}+\partial_0 B_{i\sigma}+\partial_i B_{\sigma 0}\bigg)+2\mathcal{X}B_{j0}=0 \label{eom1}\\
    &\begin{split}
        &a^{-2}\bigg[\eta^{\rho\sigma}\partial_\rho\bigg(\partial_\sigma B_{ij}
+\partial_i B_{j\sigma}+\partial_j B_{\sigma i}\bigg)
-\mathcal{H}\bigg(\partial_0 B_{ij}+\partial_i B_{j0}+\partial_j B_{0i}\bigg)\bigg]+2\mathcal{Y}B_{ij}=0
\end{split} \label{eom2}
\end{eqnarray}
Equations (\ref{constrain1}),(\ref{constrain2}),(\ref{eom1}) and
(\ref{eom2}) determine the complete evolution of the $B$ field.
\subsection{Solving for the 'electric' component}
For the rest of our discussion we will only consider the
'electric' component of the $B$ field: $B_{0i}\equiv E_i$, where
it is understood that $E$ is a vector. We don't study the
'magnetic' component, since it turns out that it is completely
regular and thus for our purposes not very interesting. If we
assume $\mathcal{X}\neq0$, then we easily find, using
(\ref{constrain1}), (\ref{constrain2}) and (\ref{eom1}) that
\begin{equation}
    a^{-2}\bigg(\eta^{rs}\partial_r\partial_s E+\partial_0\big(\mathcal{Y}^{-1}\partial_0(\mathcal{X}E)\big)\bigg)-2\mathcal{X}E=0.
\end{equation}
To study the behavior of this equation, we first rescale $E$:
\begin{equation} \label{transformation}
    {E}=\frac{\sqrt{\mathcal{Y}}}{\mathcal{X}}\tilde E
\end{equation}
to obtain
\begin{equation} \label{finaleom}
\bigg[\partial_0\partial_0-\frac{\mathcal{Y}}{\mathcal{X}}\delta^{ij}\partial_i\partial_j+M^2_{eff}\bigg]\tilde{E}=0,
\end{equation}
where the effective mass term is given by
\begin{equation}\label{effmass}
M^2_{eff}=-2\mathcal{Y}a^2+\frac{\mathcal{Y}''}{2\mathcal{Y}}-\frac{3(\mathcal{Y}')^2}{4\mathcal{Y}^2}
\,.
\end{equation}
So we see from (\ref{finaleom}) that $\tilde{E}$ behaves just as a
massive vectorfield, \emph{as long as,
${\mathcal{Y}}/{\mathcal{X}}>0$}. We see for example that in
Fourier space we have
\begin{equation}
\bigg[\partial_0\partial_0+\frac{\mathcal{Y}}{\mathcal{X}}k^2+M^2_{eff}\bigg]\tilde{E}_F=0,
\end{equation}
where $k$ is the momentum. Therefore, if
$\mathcal{Y}/\mathcal{X}>0$, $\tilde{E}$ is a fluctuating,
finite field and large values of $k$ will be suppressed. On the
other hand, if $\mathcal{Y}/\mathcal{X}<0$, for sufficiently
large momenta, $(\mathcal{Y}/\mathcal{X})k^2+M_{eff}<0$. For these momenta
the field $\tilde{E}$ grows exponentially without bounds.
Such a situation is unstable and thus not physically viable.
Therefore, if in such a case the transformation (\ref{transformation})
is regular, the theory is unphysical. One could worry as well about the cases
when, $M^2_{eff}<0$. However on dimensional grounds, the effective mass
squared scales in the worst case as, ${1}/{\eta^2}$. Such a scaling results
in a standard power-law enhancement on super-Hubble
scales~\cite{Prokopec:2005fb} and presents no problem.

\subsection{de Sitter inflation}

During de Sitter inflation (see
table \ref{tabelletje}), Eq.~(\ref{finaleom}) becomes (we use
subscripts $I$, $r$ or $m$ for the fields to indicate wether we are
considering inflation, radiation or matter era)
\begin{equation}
\big[\partial_0\partial_0-\delta^{ij}\partial_i\partial_j+M^2_{I}\;\big]
           \tilde{E}_I=0,
\end{equation}
where
\begin{equation}
 M^2_I=\frac{1}{\eta^2}\bigg((4\gamma-12\alpha-48\beta)+\frac{m^2}{H_I^2}
        \bigg).
\end{equation}
So we see that during de Sitter inflation, the field is
perfectly regular. Notice that any 'negative mass' contribution
indeed scales as ${1}/{\eta^2}$.

During de Sitter inflation
the rescaling (\ref{transformation}) is nonsingular as long as
\begin{equation} \label{constrinfl}
6\alpha+ 24\beta- 2\gamma - \frac{m^2}{2H_I^2}\neq 0
\end{equation}

\subsection{Radiation era}

During radiation era we obtain
\begin{equation}
 \bigg[\partial_0\partial_0
    -  \frac{H_I^2m^2\eta^4+4(\gamma-\alpha)}
           {H_I^2m^2\eta^4-4(\gamma-\alpha)}\delta^{ij}\partial_i\partial_j
    +  M^2_{r}
 \bigg]\tilde{E}_r=0.
\label{EOM:radiation era}
\end{equation}
Here $M_{r}$ is the effective mass during radiation, which is
given by a complicated expression. Fortunately its precise form
is not important for us. It is sufficient to say that it indeed
scales as ${1}/{\eta^2}$ and therefore it does not cause any
problems. We see however, that we might have problems in this case
with the sign of the coefficient in front of the spatial
derivatives. For example if we look at the beginning of radiation era
($\eta={1}/{H_I}$) we see that if we want
${\mathcal{Y}}/{\mathcal{X}}$ to be positive, we need that
${m^2}/{H_I^2}$ is \emph{at least},
$\mathcal{O}(\alpha-\gamma)$. In other words we approximately
need:
\begin{equation} \label{error}
m\geq |\alpha-\gamma|H_I\sim |\alpha-\gamma|\times 10^{13}~{\rm GeV}
\,,
\end{equation}
which, unless $|\alpha-\gamma|$ is very small,
contradicts Eq.~(\ref{assumption}). Therefore if we require
${\mathcal{Y}}/{\mathcal{X}}$ to be positive, we either need
to fine-tune $\alpha$ or $\gamma$ such that $\alpha-\gamma$ is sufficiently
small to satisfy the bound~(\ref{error}),
or we could use the more natural requirement that $\alpha=\gamma$.
 The latter case will be examined in more detail in
section~\ref{secalphagamma}.

 For now we assume that $|\alpha-\gamma|$ is not unnaturally small,
so we can effectively neglect the term, $H_I^2m^2\eta^4$ in
Eq~(\ref{EOM:radiation era}) at the beginning of radiation era
and we get, ${\mathcal{Y}}{/\mathcal{X}}=-1$.
Therefore we obtain an unstable equation of motion for $\tilde{E}$, implying
that $\tilde{E}$ grows exponentially during radiation era. Of course
this growth is a problem only if the transformation
(\ref{transformation}) is regular. One readily checks that during radiation
era, Eq.~(\ref{transformation}) is non-singular if
\begin{equation}
4(\alpha-\gamma)+H_I^2m^2\eta^4\neq 0.
\end{equation}
This is clearly the case, since we consider $\alpha-\gamma$ to be
dominant and nonzero. Therefore we conclude that, unless
$\alpha-\gamma$ is tiny, the equation of motion~(\ref{EOM:radiation era})
develops an instability at the beginning of radiation era.

\subsection{Matter era}

 During matter era the equation of motion becomes
\begin{equation}
\bigg[\partial_0\partial_0-\frac{H_I^2m^2(\eta_e+\eta)^6\eta_e^{-4}-128(3\alpha+6\beta-2\gamma)}{H_I^2m^2(\eta_e+\eta)^6\eta_e^{-4}-128(6\beta+\gamma)}\delta^{ij}\partial_i\partial_j+M^2_{m}\bigg]\tilde{E}_m=0,
\end{equation}
where $M_{m}$ is again a complicated -- but for our purposes not
very interesting -- expression. The factor in front of the spatial
derivatives does not cause any problems, since the term
$H_I^2m^2(\eta_e+\eta)^6\eta^{-4}$ dominates over the term
containing the $\alpha$, $\beta$ and $\gamma$.  The field
rescaling (\ref{transformation}) is during matter era non-singular if
\begin{equation}
-768\beta- 128\gamma+ H_I^2 m^2\frac{(\eta_e+ \eta)^6}{\eta^{4}}\neq 0.
\end{equation}
This relation always holds, since the last term dominates.

\subsection{Special cases}
\label{secalphagamma}

The first special case we consider is $\alpha=\gamma$. As was
noted above, the case $\alpha=\gamma$ is special, since now the
instability at the beginning of radiation era appears to be solved. We
notice that if $\alpha=\gamma$ we have
\begin{equation}
\mathcal{X}=\mathcal{Y}=a^{-2}\bigg((12\beta+2\gamma)(\mathcal{H}^2+\mathcal{H}')-\frac{a^2m^2}{2}\bigg)
\end{equation}
Now if we assume (\ref{constrinfl}) to hold ($\beta\neq
\frac{\gamma}{12}(\frac{m^2}{4H_I^2}-2\alpha)$), the only era
where we have potentially problems is radiation. However during
radiation era we find that
\begin{equation}
    \mathcal{H}^2_r+\mathcal{H}'_r=0
    \quad\quad\rightarrow\quad\quad\mathcal{X}=-\frac{m^2}{2}
\end{equation}
and therefore the equation of motion (\ref{finaleom}) reduces to
\begin{equation}
\Big[\partial_0\partial_0-\delta^{ij}\partial_i\partial_j+m^2a^2\Big]E_r=0
\quad\quad\quad(\alpha=\gamma).
\end{equation}
So we find that the unrescaled field behaves perfectly regularly
during radiation era.

 The next case we consider is the (local) situation where
$\alpha=\gamma$, $\mathcal{X}\rightarrow 0$ and $\mathcal{X}'\neq
0$. This exactly corresponds to the cases where
Eq.~(\ref{transformation}) becomes singular. Also notice that
$\mathcal{X}\rightarrow 0$ is only possible if
$\beta>-{\gamma}/{6}$. In this case one finds that
(\ref{constrain2}) reduces to
\begin{equation} \label{constraint3}
(12\beta+2\alpha)(\mathcal{H}''-\mathcal{H}^3)E=0
\end{equation}
and thus $E$ is constrained to become zero at one point in time.
Such a constraint on a dynamical degree of freedom would normally be
unacceptable. Here this is not the case however, since we also
find from (\ref{effmass}) that at that moment,
\begin{equation}
    M_{eff}\rightarrow\infty.
\end{equation}
Therefore $E$ is dynamically pushed to zero (it becomes nondynamical)
at the moment under consideration and the constraint (\ref{constraint3})
presents no problem.

 The last special case we consider is the case where
not only $\alpha=\gamma$, but also $\beta=-{\gamma}/{6}$. Now
we find that
\begin{equation}
\mathcal{X}=\mathcal{Y}=-\frac{m^2}{2}
\end{equation}
and therefore during all eras the equation of motion reduces to the
following simple form:
\begin{equation}
\Big[\partial_0\partial_0-\delta^{ij}\partial_i\partial_j+m^2a^2\Big]E=0
\quad\quad\quad
  \Big(\alpha=\gamma\quad;\quad\beta=-\frac{\gamma}{6}\Big)
.
\end{equation}
Despite the elegant form, there is unfortunately nothing that
tells us to constrain $\beta$.

\section{Evolution in a Schwarzschild background} \label{sschwarz}
The other case we will consider is the behavior of the $B$-field
in a Schwarzschild background. Although the background is clearly
spherically symmetric, there is no need for the $B$-field to also
be spherically symmetric, so we will not explicitly make it so. The
metric (in spherical coordinates) is given by
\begin{equation}
G_{\mu\nu}=\text{diag}
  \bigg\{1-\frac{2M_s}{r},-\Big(1-\frac{2M_s}{r}\Big)^{-1},
       -r^2,-r^2\sin^2\theta
  \bigg\},
\end{equation}
where $M_s$ is the Schwarzschild mass parameter. The equation of
motion for the $B$-field reduces in this spacetime to
\begin{equation}\label{schweom}
\frac{1}{2}\partial_\rho(\sqrt{-G}H^{\rho\mu\nu})+\sqrt{-G}\bigg(\frac{m^2}{2}B^{\mu\nu}-2\gamma
B^{\alpha\beta}R^\mu{}_\alpha{}^\nu{}_\beta\bigg)=0,
\end{equation}
with
\begin{equation}
\sqrt{-G}=r^2\sin{\theta}
\end{equation}
and the relevant components of the Riemann tensor are given in
appendix \ref{ageom}. Notice that (\ref{schweom}) is independent
of $\alpha$ and $\beta$ since the metric is Ricci-flat. We next
act with $\partial_\mu$ on (\ref{schweom}) and divide by
$\sqrt{-G}$ to obtain the following consistency equation:
\begin{equation}
\begin{split}
&\frac{m^2}{2}\partial_\mu B^{\mu\nu}-2\gamma\partial_\mu(B^{\alpha\beta}R^\mu{}_\alpha{}^\nu{}_\beta)\\
&+\frac{m^2}{r}B^{1\nu}-\frac{4\gamma}{r}B^{\alpha\beta}R^1{}_\alpha{}^\nu{}_\beta+\frac{m^2}{2}\cot\theta
B^{2\nu}-2\gamma\cot\theta
B^{\alpha\beta}R^2{}_\alpha{}^\nu{}_\beta =0
\end{split}
\end{equation}
The precise structure of these constraints is not very clear, so
we consider the four different components separately. We obtain
(the index $a$ in Eqs.~(\ref{con1}--\ref{con4})
indicates the angular directions: $a=2\equiv\theta,a=3\equiv\phi=3$),
\begin{eqnarray}
\mathcal{T}B^{10}+\mathcal{R}\cot\theta
B^{20}+\mathcal{S}\partial_1B^{10}+\mathcal{R}\partial_a
B^{a0}&=0\\ \label{con1} \mathcal{R}\cot\theta
B^{21}-\mathcal{S}\partial_0 B^{10}+\mathcal{R}\partial_a
B^{a1}&=0\\\label{con2} \mathcal{Q}
B^{12}-\mathcal{R}\partial_0B^{20}-\mathcal{R}\partial_1B^{21}-\mathcal{S}\partial_3B^{23}&=0\\\label{con3}
\mathcal{Q} B^{13}+\mathcal{S}\cot\theta
  B^{23}-\mathcal{R}\partial_0B^{30}
    -\mathcal{R}\partial_1B^{31}-\mathcal{S}\partial_2B^{32}&=0
\,,
\label{con4}
\end{eqnarray}
where we have defined
\begin{eqnarray}
    \mathcal{Q}&\equiv\frac{m^2}{r}+\frac{2\gamma M_s}{r^4} \label{q}\\
    \mathcal{R}&\equiv\frac{m^2}{2}-\frac{2\gamma M_s}{r^3} \label{r}\\
    \mathcal{S}&\equiv\frac{m^2}{2}+\frac{4\gamma M_s}{r^3} \label{s}\\
    \mathcal{T}&\equiv\frac{m^2}{r}-\frac{4\gamma M_s}{r^4}
\,.
\end{eqnarray}
After a rescaling
\begin{equation}\label{rescale}
B^{10}\rightarrow\tilde{B}^{10}\frac{\mathcal{R}}{\mathcal{S}}\quad\quad\quad
B^{23}\rightarrow\tilde{B}^{23}\frac{\mathcal{R}}{\mathcal{S}},
\end{equation}
we can combine the constraints in the following more elegant form:
\begin{equation}\label{schwcon}
\partial_\mu \tilde{B}^{\mu\nu}
 +\frac{\mathcal{Q}}{\mathcal{R}}\tilde{B}^{1\nu}
 +\cot\theta\tilde{B}^{2\nu} =0
.
\end{equation}
Next we consider the (unrescaled) equations of motion for the
$B$-field. Using (\ref{schweom}) we find that
\begin{equation} \label{schweom01}
\partial_\rho H^{\rho\mu\nu}+\frac{2}{r}H^{1\mu\nu}+\cot\theta H^{2\mu\nu}+4\delta^\mu_{[0}\delta^\nu_{1]}(\mathcal{S}-\mathcal{R})B^{01}+4\delta^\mu_{[2}\delta^\nu_{3]}(\mathcal{S}-\mathcal{R})B^{23}+2\mathcal{R}B^{\mu\nu}=0
\end{equation}
We now focus on $B^{01}$. After the rescaling (\ref{rescale}), we
obtain from the 01-component of (\ref{schweom01}):
\begin{equation} \label{thing}
\begin{split}
&\frac{\mathcal{R}}{\mathcal{S}}\bigg(\cot\theta\partial^2+\partial^2\partial_2+\partial^3\partial_3+2\mathcal{S}\bigg)\tilde{B}^{01}\\
&+\cot\theta(\partial^0\tilde{B}^{12})+\partial^0(\partial_2\tilde{B}^{12}+\partial_3\tilde{B}^{13})\\
&+\cot\theta(\partial^1\tilde{B}^{20})+\partial^1(\partial_2\tilde{B}^{20}+\partial_3\tilde{B}^{30})=0
\end{split}
\end{equation}
Next we use (\ref{schwcon}) to write the last two lines in terms
of $\tilde{B}^{01}$ to obtain
\begin{equation} \label{tempeom}
\bigg[\partial^0\partial_0+\partial^1\partial_1+\frac{\mathcal{R}}{\mathcal{S}}\bigg(\partial^2\partial_2+\cot\theta\partial^2+\partial^3\partial_3\bigg)+2\mathcal{R}+\partial^1\Big(\frac{\mathcal{Q}}{\mathcal{R}}\Big)+\frac{\mathcal{Q}}{\mathcal{R}}\partial^1\bigg]\tilde{B}^{01}=0
\end{equation}
We rescale the field to get rid of the single derivative term.
\begin{equation}
\hat{B}^{01}=\lambda\tilde{B}^{01}=-\frac{\sqrt{r}\sqrt{4\gamma
M_s-m^2 r^3}}{8\gamma M_s+m^2 r^3}
\end{equation}
where we have defined
\begin{equation}
\lambda=\frac{\sqrt{r}}{\sqrt{4\gamma M_s-m^2 r^3}}.
\end{equation}

 Finally we divide (\ref{tempeom}) by $g^{00}$ and
get
\begin{equation}
\label{eom01}
\bigg[\partial_0\partial_0-\frac{(r-2M_s)^2}{r^2}\partial_1\partial_1-\frac{r-2M_s}{r}\frac{\mathcal{R}}{\mathcal{S}}L^2+M^2_{eff}\bigg]\hat{B}^{01}=0,
\end{equation}
where the effective mass term is now given by
\begin{equation}
    M_{eff}^2=\bigg(1-\frac{2M_s}{r}\bigg)\bigg(2\mathcal{R}-g^{11}\partial_1\partial_1\lambda\bigg).
\end{equation}
and the operator $L^2$ is defined to be
\begin{equation} \label{L2}
L^2=\frac{1}{r^2}\bigg(\partial_2\partial_2+\cot\theta\partial_2+\frac{1}{\sin^2\theta}\partial_3\partial_3\bigg)
\end{equation}
When the solution of Eq.~(\ref{eom01}) is written in a factorized
form, $\tilde{B}^{01}=\sum_{lm}b_{lm}(r)Y_{lm}(\theta,\phi)$, the
operator $L^2$ generates a centrifugal barrier term,
$-{l(l+1)}/{r^2}$, where $l=0,1,2,..$ is the multipole moment.
Therefore we see that, depending on the sign of
${\mathcal{R}}/{\mathcal{S}}$, we have similar problems as with
the cosmological solutions (\ref{finaleom}). If
${\mathcal{R}}/{\mathcal{S}}$ is positive (we only consider fields
outside the Schwarzschild radius, $r>2M_s$), we have a normal,
well behaving field, since the field for high values of $l$ is
suppressed. On the other hand if $\mathcal{R}/\mathcal{S}$ is
negative, high values of $l$ are no longer suppressed, and the
field $\tilde{B}^{01}$ grows exponentially without a limit. Such a
situation is clearly unstable. We evaluate
${\mathcal{R}}/{\mathcal{S}}$ using (\ref{r}) and (\ref{s}) and
find that
\begin{equation} \label{rs}
\frac{\mathcal{R}}{\mathcal{S}}=\frac{m^2r^3-4\gamma
M_s}{m^2r^3+2\gamma M_s}.
\end{equation}
We see that there are certainly finite values for $r$ and $M_s$
where this quantity becomes negative. In these regions the field
is unstable, as was explained above.
Note that ${\cal R}/{\cal S}$ is negative when,
\begin{equation}
 r<r_{\rm cr} = \begin{cases}\Big(\frac{4\gamma M_s}{m^2}\Big)^\frac13
                                     &{\rm for}\quad \gamma>0 \cr
                       \Big(\frac{-2\gamma M_s}{m^2}\Big)^\frac13
                                     &{\rm for}\quad \gamma<0
                \end{cases}
\,.
\label{Schwarz:rcr}
\end{equation}
Since there is no reason to believe that $\gamma$ is particularly
small and that $m^{-1}$ should be less than the Schwarzschild
radius, $r_s =2 M_s$, for \emph{all} spherically symmetric mass distributions
in the Universe,
we conclude that the condition~(\ref{Schwarz:rcr}) will be
satisfied at least somewhere in the Universe.

 Since the $B$-field should be stable for all values of $M_s$ and $r$,
for a sensible theory the sign of
${\mathcal{R}}/{\mathcal{S}}$ should be independent of $M_s$ and $r$.
The only solution for this requirement is fixing
$\gamma=0$. In this case we have
${\mathcal{R}}/{\mathcal{S}}=1$ and the equation of motion
(\ref{eom01}) simplifies to
\begin{equation}
\bigg[\partial_0\partial_0
    -\frac{(r-2M_s)}{r}
         \bigg(\frac{(r-2M_s)}{r}\partial_1\partial_1+L^2\bigg)+M_{eff}^2
         \bigg)
 \bigg]\tilde{B}^{01}=0\quad\quad\quad\quad(\gamma=0),
\end{equation}
where
\begin{equation}
M_{eff}^2=\frac{(r-2M_s)}{r}\bigg(m^2+\frac{2r-4M_s}{r^3}\bigg)\quad\quad\quad\quad(\gamma=0).
\end{equation}
Thus we see that (outside the Schwarzschild radius) the mass gets
a positive enhancement. The field behaves perfectly regular and
similar --but not equal-- to a 'normal' massive vectorfield in a
Schwarzschild geometry.

\section{Analysis}
 \label{sdisc}

\subsection{Discussion of the evolution in a FRLW universe}

The calculations done in section~\ref{scosmo} lead to the
conclusion that, if we assume that the mass of the $B$-field is
much less then $H_I$ and we want to avoid fine tuning, the couplings
of the $B$-field to the background curvature, we must choose
$\alpha=\gamma$. If we do not make this choice, then
at the beginning of radiation era the $B$ field becomes unstable
and grows exponentially without a bound. We have also shown that the field
equations are regular for this choice of the parameters. Looking
back at the calculation we see that the reason these instabilities
occur lies in the fact that there are two functions in front of
the different components in the field equations ($\mathcal{X}$ and
$\mathcal{Y}$, see Eqs.~(\ref{X}--\ref{Y})), that could have a relative
minus sign. This then leads to the possibility of the 'wrong' sign
in front of the spatial derivatives~(\ref{eom01}). Indeed the choice
$\alpha=\gamma$ is the only possibility where
$\mathcal{X}=\mathcal{Y}$ independent of the free parameters of
the theory.

 The fact that we have a constraint on the couplings
to the background, means that given a non-linearized lagrangian,
one of the parameters of the decomposition (\ref{decomposition})
is actually fixed. So, while we are not able to a priori tell how
to decompose the metric into its symmetric and antisymmetric
components, consistency of the theory during all epochs of the
Universe gives us one constraint. Since there are two arbitrary
parameters in (\ref{decomposition}) we are however not able to
completely fix the decomposition. The special choice
$\beta=-{\gamma}/{6}$ for example would completely fix it,
but except perhaps for simplicity, we are not aware of any physical reason
why would this be the correct choice.

 One could hope that looking at curved FLRW spacetimes might give a
new constraint. These cases have also been checked and lead to
essentially the same results.

\subsection{Discussion of the evolution in a Schwarzschild background}
\label{sdiscschwarz}

 In section~\ref{sschwarz} we considered the evolution of a
non-constrained $B$-field in a spherically symmetric
(Schwarzschild) background. We found that, unless we choose
$\gamma=0$, the $B$-field always has an unstable mode for
particular values of $M_s$ and $r$. Since there is no reason to
exclude these values, we can only conclude that we should take
$\gamma=0$ for the theory to be consistent. The seeds for these
inconsistencies lie, similarly as in the cosmological case, in the
possibility of a relative minus sign between the coefficients of
the different components in the field equations (in particular the
coefficients $\mathcal{R}$ and $\mathcal{S}$, see
(\ref{r}--\ref{s})). Unfortunately, as we have shown in appendix
\ref{alin}, $\gamma=0$ is not possible within NGT in first
order formalism. In fact one finds that an initial lagrangian that
linearizes to a theory with $\gamma=0$ also has the coefficient in
front of $H^2$ equal to zero. So this becomes a trivial,
nondynamical theory.

One might hope that perhaps a formulation of the theory in the
second order formalism solves the problems and indeed one finds
that now it is possible to find a full Lagrangian that gives
$\gamma=0$. However in~\cite{Damour:1992bt} it is shown that in
such a case, one always gets at some higher order $n$ in $B$
terms of the form $R B^n$. In the next section~\ref{ssextend} we
argue that these terms play exactly the same
r$\hat{\text{o}}$le as the $RB^2$ terms we considered.

 Therefore the instabilities are present both in the first and
second order formulations. We must thus conclude that there is no
geometric full Lagrangian, as considered in appendix~\ref{alin},
that correspond to a stable theory at linearized level.

This result is no big surprise, since it was already noted by
Clayton in ~\cite{Clayton:1996dz} that NGT has linearization
instabilities. The nature of the instabilities we describe is
however essentially different from the Cauchy instabilities
Clayton finds. There are two proposed solutions for the
instabilities Clayton discusses. First of all one might hope that
corrections beyond linearized level might stabilize the theory. We
address this issue in the next section where we are lead to the
conjecture that our instabilities make NGT unstable beyond
linearized level. Secondly, it was proposed in~\cite{Moffat:1996hf}
to modify the theory with lagrange
multipliers that dynamically put the unstable modes to zero. In
section~\ref{sdynamically} we show that these dynamical
constraints do not resolve the instabilities we address.

\subsection{Extending the results beyond linearized level: A conjecture}
\label{ssextend}

Our results so far only apply to the
linearized version of NGT. Since we showed that in this limit
the field grows without a bound, one is lead to the conclusion that
the linearization {\it Ansatz} of a small $B$-field, is not anymore
valid. We therefore need to consider NGT at higher orders. Already
if we look at the quadratic order, we encounter many mathematical
complexities, since (as is clear from (\ref{lagrangian}))
Einstein's field equations get a modification of order $B^2$.
Therefore the standard GR solutions $G_{\mu\nu}$ for the symmetric
metric are not anymore a solution of the theory beyond linear
level. Instead we assume we have (at quadratic order)
\begin{equation} \label{f}
g_{(\mu\nu)}=G_{\mu\nu}+f_{\mu\nu}(B^2,H^2,G),
\end{equation}
where $f_{\mu\nu}$ is some unknown function of order $B^2$ that vanishes if
$B^2 \rightarrow 0$. Therefore the $B$-field backreacts on the
background metric. It has been argued by Clayton~\cite{Clayton:1996dz}
that such a backreaction might stabilize
the theory, since the $B$-field in such a case acquires additional
degrees of freedom (similarly to~\cite{Isenberg:1977}).

 We will now look in a more detailed manner at the equations of
motion for the $B$-field at higher order in the specific case
where the GR background is the Schwarzschild metric. First of all
notice that there will not be a quadratic contribution to the
$B$-field equations of motion. The reason is that there will be no
cubic contributions to the lagrangian, just as there is no linear
contribution. So therefore the first order we can consider is
quartic in the lagrangian and therefore cubic in the field
equations. Although the function $f_{\mu\nu}$ may be complicated,
it is clear that a typical term that arises in these equations is
($R_{\rho\alpha\sigma\beta}$ refers to the GR solution),
\begin{equation}
R_{\rho\alpha\sigma\beta}B^{\rho\sigma}B^{\alpha\beta}B^{\mu\nu}.
\end{equation}
An explicit calculation shows that this term is equal to
\begin{equation} \label{cubic}
\begin{split}
&\bigg(\frac{2M_s}{r^3}(B^{01})^2+\frac{M_s(2M_s-r)}{r^2}(B^{02})^2+(B^{03})^2\frac{M_s(2M_s-r)}{r^2}\sin^2\theta\\&-(B^{12})^2\frac{M_s}{2M_s-r}-(B^{13})^2\frac{M_s}{2M_s-r}\sin^2\theta-(B^{23})^22M_sr\sin^2\theta\bigg)B^{\mu\nu}.
\end{split}
\end{equation}
Therefore we see that the different components of the $B$ field
appear with different coefficients in front of them. More
importantly we see that, depending on the values of $M_s$ and $r$,
there could be a relative minus sign between various terms.
Looking back at the calculations in section~\ref{sschwarz}, one
sees that these relative minus signs are also present at linear
level (although the coefficients differ, see
Eq.~(\ref{schweom01})). We also have seen that, depending on
whether we have such a relative minus sign, we may or may not get
a relative minus sign between the $\partial_0^2$ and the
$\partial_i^2$ terms, and thus we may or may not get
instabilities. The question now is if the relative minus signs in
(\ref{cubic}) implies the same for the cubic sector of the theory.

 First we assume we can solve the cubic equations of motion
for only one of the components of the $B$-field (which we call
$b$). Then, in analogy with section~\ref{sschwarz}, we obtain an
equation of motion of the following form (we assume that single
derivative terms can be eliminated by a rescaling of $b$)
\begin{equation} \label{diff}
(1+\alpha_1 b^2)\partial_0^2 b +\alpha_2 b+\alpha_3
b^3+\alpha_4b(\partial_0 b)^2=0.
\end{equation}
The precise value of the various $\alpha_i$'s depends on the
specific form of the higher order lagrangian and the function
$f_{\mu\nu}$ in Eq. (\ref{f}). $\alpha_2$ and $\alpha_3$ depend
also on the momenta (spatial derivatives) of the field
component. Eq.~(\ref{cubic}) strongly
suggests that the sign of the different $\alpha$'s
is dependent on the values of $M_s$ and $r$. Therefore we see that
$\alpha_2$ and/or $\alpha_3$ can be negative. Since
$|\alpha_{2,3}|$ can be as large as we want (because of the
momentum dependence), we find by numerical analysis that
(\ref{diff}) always can develop instabilities (independent of the
sign of $\alpha_1$ or $\alpha_4$). Therefore, although we have not
proven it, the similarity between (\ref{cubic}) and
(\ref{schweom01}) indicates that, for certain choices
of the parameters and coordinates, the necessary conditions
for the instabilities to occur can always be met. Thus it appears that
NGT is also unstable at cubic order. In fact since at higher
orders in $B$ there is no new structure to be expected, we are
lead to the conjecture that NGT is unstable at all orders.

The largest pitfall in the above argument is that it might not be
possible to decouple the different modes of the $B$-field. In such
a case (\ref{diff}) would change to an equation of the form
\begin{equation}
(1+\alpha_1 b_1^2)\partial_0^2 b_1 +\alpha_2 b_1+\alpha_3
b_1^3+\alpha_4b_1(\partial_0 b_1)^2={\cal D}[b_i]
\,,
\label{cubic B equation}
\end{equation}
where ${\cal D}$ is some second order differential operator acting in
general on all components $b_i\neq b_1$
and with derivatives with respect to -- possibly
-- all four coordinates. It is clear from the above discussion
that the 'homogeneous' differential equation (with ${\cal D}=0$) would
give instabilities.
The coupling between the differential equations induced by
the operator ${\cal D}$ in Eq.~(\ref{cubic B equation})
may be such that the field components mutually stabilize each other.
The question is whether this indeed
can happen in a physically meaningful way. To investigate this
consider the following toy model:
\begin{equation}
 \label{toy}
\begin{split}
    f''(t)-k_f^2f(t)=\alpha g'(t)\\
    g''(t)-k_g^2 g(t)=\beta f'(t).
\end{split}
\end{equation}
We expect this system to be representative of the dynamics of
small $B$-field fluctuations around some general background metric
which also includes nonlinear antisymmetric metric contributions.
In Eq.~(\ref{toy}) a {\it prime} denotes a time derivative, $k_f$
and $k_g$ refer to the momenta of the field components $f$ and
$g$, $\alpha$ and $\beta$ are functions of the background metric
fields '$G_{\mu\nu}$' and '$B_{\mu\nu}$', possibly their first
space or time derivatives (and thus they may be linear in $k_f$
and $k_g$), and the parameters of the theory ($M_s$ and $r$ in the
Schwarzschild case). In general the structure of the second
spatial derivative terms in Eq.~(\ref{toy}) is not so simple ({\it
cf.} Eq.~(\ref{eom01})), but the case considered here is
representative for a more general situation. Indeed, the above
discussion shows that the general second order spatial derivative
terms reduce to the form~(\ref{toy}) for certain ranges of the
coordinates. Looking at~(\ref{thing}) shows that the coupling in
Eq.~(\ref{toy}) is exactly the type of coupling we would expect
between the different field components in NGT after we rescale
away all single derivative terms. A more general coupling could
also include $\rho g''(t)+\sigma g(t)$, but this would not alter
the following discussion.
The system (\ref{toy}) gives rise to
the following eigenvalue equation
\begin{equation} \label{eigenvalue}
(\omega^2+k_f^2)(\omega^2+k_g^2)+\omega^2\alpha\beta=0,
\end{equation}
when the solution is decomposed as
\begin{equation}
\begin{split}
    f(t)=\sum_{i=1}^4 a_i e^{i\omega_i t}\\
    g(t)=\sum_{i=1}^4 b_i e^{i\omega_i t}
\,.
\end{split}
\end{equation}
Thus the system is unstable if at least for one root
$\Im(\omega_i)<0$ ($\Im$ indicates the imaginary part). We can
solve (\ref{eigenvalue}) to obtain
\begin{equation}
\omega^2=\frac{1}{2}\bigg(-k_f^2-k_g^2-\alpha\beta
        \pm\sqrt{(k_f^2+k_g^2+\alpha\beta)^2-4k_f^2k_g^2}\bigg).
\label{roots}
\end{equation}
First note that if $\Im(\omega^2)\neq 0$, then there is at least
one root for which $\Im(\omega)< 0$ is satisfied, and the system
is unstable. So the necessary condition for the stability of
the system is, $\Im(\omega^2) = 0$.  From Eq.~(\ref{roots})
we then easily infer that this condition is obeyed
provided,
\begin{equation}
   \alpha\beta>-(k_f-k_g)^2
\end{equation}
or
\begin{equation}
   \alpha\beta<-(k_f+k_g)^2.
\end{equation}
It is clear that it is impossible to satisfy the second equation
for all values of the momenta. However the first equation can be
satisfied if $\alpha\beta>0$. When $\Im(\omega^2) = 0$ is
satisfied, the sufficient condition for stability is then,
$\Re(\omega^2)>0$ (indicating the real part), which implies,
\begin{equation}
 k_f^2+k_g^2+\alpha\beta < 0
\,,\qquad (\alpha\beta>0)
\end{equation}
Since we also require, $\alpha\beta>0$, both are impossible to
satisfy simultaneously.
We hence conclude that there always is an unstable root
of Eq.~(\ref{roots}) with $\Im(\omega)<0$, which destabilizes the
system~(\ref{toy}).

While the full analysis of cubic, coupled differential equations
is much more difficult, based on the analysis of small
fluctuations around a nonlinear background, we have argued that
the same results apply: it is not possible to couple the unstable
modes in such a way that the system stabilizes for all values of
the momenta. The fact that the coupling constants in general
depend on $M_s$ and $r$ only make matters worse.

 Finally, our analysis of the case
where the different components of the $B$-field
mutually couple, indicates that our conjecture holds
and NGT is indeed unstable beyond linearized level.

\subsection{Dynamically constrained NGT}
\label{sdynamically}

 In order to solve the instabilities discovered in NGT by
Clayton~\cite{Clayton:1996dz}, Moffat proposed the introduction of an
extra term to the full Lagrangian~\cite{Moffat:1996hf} that
looks like
\begin{equation}
  \mathcal{L}_{DNGT}=\sqrt{-g}g^{\mu\nu}J_{[\mu}\phi_{\nu]}.
\end{equation}
Here $J_\mu$ is some source vector and $\phi_\nu$ plays the
r$\hat{\text{o}}$le of a Lagrange multiplier. This new term then
leads to the dynamical constraint
\begin{equation}
g^{[i0]}=0.
\end{equation}
Since these are exactly the unstable modes Clayton found, these
instabilities are dynamically resolved. One might hope that this
extra constraint also solves the instabilities we have found.
\subsubsection{The cosmological case}
For the cosmological case we find from (\ref{constrain1}),
(\ref{constrain2}), (\ref{eom1}) and (\ref{eom2}) that (if
$\mathcal{Y}\neq 0 $) the dynamics of the dynamically constrained
theory are determined by
\begin{equation}
\Big[\partial_0\partial_0-\delta^{rs}\partial_r\partial_s
-\mathcal{H}\partial_0 +2a^2\mathcal{Y}\Big]B_{ij}=0.
\end{equation}
After a rescaling
\begin{equation}
B_{ij}=\frac{1}{\sqrt{a}}\tilde{B_{ij}}
\end{equation}
one finds that
\begin{equation}
\Big[\partial_0\partial_0-\delta^{rs}\partial_r\partial_s
+M_{eff}^2\Big]\tilde{B}_{ij}=0,
\end{equation}
which indeed has well behaving solutions.
\subsubsection{The Schwarzschild case}
However for the Schwarzschild case we will now show that the
problems remain, although in a somewhat different disguise. We focus
on the $12$-component of the field equations (\ref{schweom}) and
obtain
\begin{equation}
\Big[\partial^0\partial_0
  +\partial^3\partial_3+2\mathcal{R}\Big]B^{12}
  +\partial_3(\partial^1B^{23}+\partial^2B^{31})=0.
\end{equation}
Using the unrescaled constraint equations (\ref{con2}) and
(\ref{con3}) we obtain
\begin{equation} \label{lalala}
\Big[\partial^0\partial_0+\frac{\mathcal{R}}{\mathcal{S}}\partial^1\partial_1
           +\partial^2\partial_2
           +\cot\theta\partial^2
           +\partial^3\partial_3
           \Big]B^{12}
           +g^{11}\Big[\Big(\partial_1\frac{\mathcal{R}}{\mathcal{S}}
           +\frac{\mathcal{Q}}{\mathcal{S}}\Big)\partial_1
           +\Big(\partial_1\frac{\mathcal{Q}}{\mathcal{S}}\Big)\Big]B^{12}+\partial^2\cot\theta
           B^{12}
           =0
.
\end{equation}
Next we rescale $B^{12}=f(r)\tilde{B}^{12}$ with some function
$f(r)$ that is determined by the requirement that the single
derivative terms in (\ref{lalala}) cancel:
\begin{equation}
\bigg(\frac{\mathcal{Q}}{\mathcal{S}}
 +\partial_1\frac{\mathcal{R}}{\mathcal{S}}\bigg)f(r)
 + 2\frac{\mathcal{R}}{\mathcal{S}}\partial_1f(r)
 = 0
.
\end{equation}
This equation can be solved, although the resulting expression is
rather complicated. The final result for the equation of motion is
\begin{equation}
\bigg[\partial_0\partial_0
     -\frac{(r-2M_s)^2}{r^2}\frac{\mathcal{R}}{\mathcal{S}}
                                            \partial_1\partial_1
    - \frac{r-2M_s}{r}L^2
    +\text{mass\;term}\bigg]\tilde{B}^{21}=0,
\end{equation}
where the operator $L^2$ is defined in (\ref{L2}) and the mass
term is some complicated function of $r$.  We clearly see that our
troubles are not resolved, since once again the sign of
$\mathcal{R}/\mathcal{S}$ determines whether the solutions are
stable. Exactly as in section \ref{sschwarz} (cf. Eq~(\ref{rs})),
there are certainly values of $M_s$ and $r$ where the
$\mathcal{R}/\mathcal{S}$ becomes negative. This results in an
unstable solution for $\tilde{B}^{21}$. It can be shown that the
function $f(r)$ is also regular in this regime.

 As a final hope for stability one might try to impose even more
Lagrange multipliers to also remove the unstable mode discussed
above. However one finds that the mode $B^{31}$ exhibits the same
behavior as $B^{21}$ and therefore should also be constrained to
zero. Then one can show that the only remaining mode, $B^{23}$ is
unstable exactly when $B^{21}=B^{31}=0$. Therefore we conclude
that it is not possible to use dynamical constraints to remove the
instabilities we found.

\section{Conclusions and outlook}
\label{scon}

We consider the evolution of the Linearized NGT lagrangian in
two different GR backgrounds. The first case we consider is a
FLRW universe during three different cosmological eras. We show
that, if the mass of the $B$-field is much smaller then $H_I$, the
field always undergoes an unbounded growth at the beginning of
radiation era. The simplest solution for this instability problem
is to fix the parameters $\alpha=\gamma$ in the lagrangian
(\ref{lagrangian}).

 The second case we consider is a Schwarzschild background.
 Here the problems are much more severe.
First of all the occurrence of instabilities does not depend on
the smallness of the mass of the $B$-field. Secondly the only way
to remove the problems is by choosing $\gamma=0$. Now this choice
is not available in the first order formalism of NGT and therefore
we conclude that the only stable linearized lagrangian must have
the form
\begin{equation} \label{finallag}
    \mathcal{L}=\sqrt{-G}\Big[
                               R+2\Lambda-\frac{1}{12}H^2
                              +\Big(\frac{1}{4}m^2 + \beta R\Big)B^2
                         \Big]
               +\mathcal{O}(B^3)
\end{equation}
and, as we show in Appendix~\ref{alin}, this form cannot be
obtained by linearizing NGT.

Since these results apply only for the linearized case, one might
hope that considering higher order corrections stabilize the
theory. One particular feature of higher order corrections is that
Einstein's field equations get modified and therefore the standard
GR solutions for the symmetric metric also get a modification of
order $B^2$. However, we have shown in Eq~(\ref{cubic}) that the
seeds for the instabilities are also present at higher order and
therefore, if we can decouple the equations of motion for each
mode, there is nothing that can prevent the instabilities from
growing. If we cannot decouple the equations of motion, the
situation is more subtle. Now in principle we could hope that two
(or more) unstable fields could couple in such a way that they
stabilize each other. We have showed however by considering a
simple, but representative example (see Eq.~(\ref{toy})), that it
is in general not possible to get rid of the instabilities. While
we do not know exactly the nonlinear field equations, this result
is most probably applicable to any coupling between the field
components one can construct, up to any order. Therefore, while
our discussion does not explicitly prove that the theory is
unstable beyond linearized level, we have gathered enough evidence
to conjecture that NGT is unstable for all orders in $B$.

One might also hope that modifying NGT could stabilize the theory.
One modification of NGT that one can consider is dynamically
constrained NGT~\cite{Moffat:1996hf}. This version of the
theory was introduced to solve the problems with linearization
instabilities discovered by Clayton~\cite{Clayton:1996dz} and
therefore it might also resolve the instabilities we consider.
We have shown that, while this modification does solve the
instabilities in the FLRW background, the evolution in a
Schwarzschild background remains unstable. We have also shown that we cannot
add extra constraints in such a way that the
unstable modes we have found disappear.

Another, perhaps more promising, way to go is to try to find
theories with a nonsymmetric metric that linearize to the form in
Eq.~(\ref{finallag}). One possibility might be to consider a
theory on a hermitian, complex 4 dimensional manifold. In such a
theory, the imaginary part of the metric must be antisymmetric and
thus plays the r$\hat{\text{o}}$le of the $B$-field. It is argued
in ~\cite{Chamseddine:2005at} that the extra symmetries on the
complex manifold, could act as diffeomorphisms for the $B$-field,
thereby saving the gauge invariance~(\ref{gauge}). Such a theory
has the properties we need for a physically viable theory,
suggesting a possible direction to continue the study of
nonsymmetric theories of gravity.

\section{Acknowledgements}

 We would like to thank Wessel Valkenburg for bringing the original
idea that lead to this paper. We would also like to thank Willem
Westra for many interesting discussions and insights on the issue
of NGT.

\appendix
\section{Linearized NGT}
 \label{alin}

In this appendix we sketch the calculation of the linearized NGT
lagrangian, starting from the most general, covariant, two
derivative lagrangian. We work within the first order formalism,
where the connection is considered to be a dynamical field and
use the following notation and conventions,
\begin{equation}
\label{definitions}
\begin{split}
g_{\mu\nu}\quad\quad&:\quad\quad \text{full, nonsymmetric
metric}\\
W^\alpha_{\mu\nu}\quad\quad&:\quad\quad \text{full, nonsymmetric
connection}\\
\mathcal{D}\quad\quad&:\quad\quad \text{covariant derivative
w.r.t. the
full connection}\\
\nabla\quad\quad&:\quad\quad \text{covariant derivative w.r.t. the
zeroth order (Levi-Civit\`a) connection}\\
Q\quad\quad&:\quad\quad \text{geometric tensors of the full theory}\\
R\quad\quad&:\quad\quad \text{geometric tensors of the zeroth order theory (GR)}\\
W_\mu\quad\quad&=\quad\quad W^\alpha_{[\mu\alpha]}\\
g_{(\mu\nu)}\quad\quad&=\quad\quad
\frac{1}{2}(g_{\mu\nu}+g_{\nu\mu})\\
g_{[\mu\nu]}\quad\quad&=\quad\quad
\frac{1}{2}(g_{\mu\nu}-g_{\nu\mu})\\
g_{\mu\alpha}g^{\mu\beta}\quad\quad&=\quad\quad g_{\alpha\mu}g^{\beta\mu}
 = \delta_\alpha^\beta \neq g_{\alpha\mu}g^{\mu\beta}\\
\frac{1}{16\pi G_N}\quad\quad&=\quad\quad 1
\end{split}
\end{equation}
and we use the following expansion of $g_{\mu\nu}$ in terms of its
symmetric ($G$) antisymmetric ($B$) part
\begin{equation}
\begin{split}
    g_{\mu\nu}&=G_{\mu\nu}+B_{\mu\nu}+\rho B_{\mu\alpha}B^\alpha{}_\nu+\sigma B^2 G_{\mu\nu} +\mathcal{O}(B^3)\\
    g^{\mu\nu}&=G^{\mu\nu}+B^{\mu\nu}+(1-\rho) B^{\mu\alpha}B_\alpha{}^\nu+\sigma B^2 G^{\mu\nu}
    +\mathcal{O}(B^3).
\end{split}
\end{equation}
This implies that
\begin{equation}
\sqrt{-g}=\sqrt{-G}\bigg(1+\frac{1}{2}\Big(\frac{1}{2}-\rho+4\sigma\Big)B^2
                   \bigg)
\end{equation}
We use $G$ to raise and lower indices. For other terms we use the
subscript $_{(n)}$ to indicate it is $n^{th}$ order in $B$ (in
general when there is no such subscript will mean a 'full' quantity). The
Riemann tensor is defined in terms of the connection
\begin{equation} \label{riemann}
Q^\mu{}_{\nu\alpha\beta}=\partial_\alpha
W^\mu_{\nu\beta}-\partial_\beta
W^\mu_{\nu\alpha}-W^\sigma_{\nu\alpha}
W^\mu_{\sigma\beta}+W^\rho_{\nu\beta} W^\mu_{\rho\alpha}
\end{equation}
and the Ricci tensor is also defined as usual
\begin{equation}
Q_{\mu\nu}\equiv R^{\lambda}{}_{\mu\lambda\nu}=\partial_\lambda
W^\lambda_{\mu\nu}-\partial_\nu W^\lambda_{\mu\lambda}-
W^\sigma_{\mu\lambda}W^\lambda_{\sigma\nu}+W^\rho_{\mu\nu}W^\tau_{\rho\tau}.
\end{equation}
Since the Riemann tensor (\ref{riemann}) does not have
its usual symmetries (in fact it is only antisymmetric in
$\alpha$ and $\beta$), we can make another, independent
contraction (that would be zero in GR)
\begin{equation}
P_{\mu\nu}\equiv R^{\lambda}{}_{\lambda\mu\nu}=\partial_\mu
W^\lambda_{\lambda\nu}-\partial_\nu W^\lambda_{\lambda\mu}
\end{equation}
 Finally we define our covariant derivatives by the
`$+-$' relation:
\begin{equation}
\nabla_\mu g_{\alpha\beta}=\partial_\mu
g_{\alpha\beta}-W^\rho_{\alpha\mu} g_{\rho\beta}-
W^\rho_{\mu\beta} g_{\alpha\rho}
\end{equation}
And we choose our Lagrangian in such a way that the other possible
definitions of the connection do not give new information. Our
starting lagrangian is~\cite{Damour:1992bt}
\begin{equation} \label{lag}
\begin{split}
\mathcal{L}=\sqrt{-g}g^{\mu\nu}\bigg[&Q_{\mu\nu}+a_1
P_{\mu\nu}+a_2\partial_{[\mu}W_{\nu]}\\
&+b_1 \mathcal{D}_\lambda W^\lambda_{[\mu\nu]}+b_2
W^\lambda_{[\mu\alpha]}W^\alpha_{[\lambda\nu]}+b_3W^\lambda_{[\mu\nu]}W_\lambda\\
&g^{\lambda\delta}g_{\alpha\beta}\bigg(c_1W^\alpha_{[\mu\lambda]}W^\beta_{[\nu\delta]}+c_2W^\alpha_{[\mu\nu]}W^\beta_{[\lambda\delta]}+c_3W^\alpha_{[\mu\delta]}W^\beta_{[\nu\lambda]}\bigg)\\
&+dW_\mu W_\nu+2\Lambda],
\end{split}
\end{equation}
where the parameters $a_1$, $a_2$, etc. are
unconstrained~\footnote{Moffat's theory has $a_1=-{1}/{2}$,
$a_2=-2$, $h=-{1}/{6}$ and all other parameters
zero~\cite{Moffat:1994hv}. A possible restriction on the
parameters is 'transposition invariance': $g_{\mu\nu}\rightarrow
g_{\nu\mu}$,
 $\Gamma^\alpha_{\mu\nu}\rightarrow \Gamma^\alpha_{\nu\mu}$, $W_\mu\rightarrow-W_\mu$. Einstein assumed that this symmetry would imply that matter and antimatter couple in the same way to gravity. Invariance under this symmetry fixes $a_1=-{1}/{2}$}
 and $\Lambda$ is the cosmological constant.

 \subsection{Calculating the connection}

To calculate the connection we vary this lagrangian with
respect to the connection and obtain
\begin{equation} \label{fullcompat}
\begin{split}
    &-\partial_\eta(\sqrt{-g}g^{\rho\sigma})+b_1\partial_\eta(\sqrt{-g}g^{[\sigma\rho]})+\delta^\sigma_\eta\partial_\nu(\sqrt{-g}g^{\rho\nu})+2a_1\delta^\rho_\eta\partial_\nu(\sqrt{-g}g^{[\sigma\nu]})\\
    &\frac{a_2}{2}\bigg(\delta^\sigma_\eta\partial_\nu(\sqrt{-g}g^{[\rho\nu]})+\delta^\rho_\eta\partial_\nu(\sqrt{-g}g^{[\nu\sigma]})\bigg)\\
    &+\sqrt{-g}\bigg\{-g^{\mu\sigma}W^\rho_{\mu\eta}-g^{\rho\nu}W^\sigma_{\eta\nu}+g^{\mu\nu}\delta^\sigma_\eta W^\rho_{\mu\nu}+g^{\rho\sigma}W^\lambda_{\eta\lambda}\\
    &+b_1\bigg(-g^{[\mu\sigma]}W^\rho_{\mu\eta}-g^{[\rho\nu]}W^\sigma_{\eta\nu}+g^{[\mu\nu]}\delta^\sigma_\eta W^\rho_{\mu\nu}+g^{[\rho\sigma]}W^\lambda_{\eta\lambda}\bigg)\\
    &+\frac{b_2}{2}\bigg(g^{\mu\sigma}W^\rho_{[\mu\eta]}+g^{\rho\nu}W^\sigma_{[\eta\nu]}-g^{\sigma\nu}W^\rho_{[\eta\nu]}-g^{\mu\rho}W^\sigma_{[\mu\eta]}\bigg)\\
    &+b_3\bigg(g^{[\rho\sigma]}W_\eta+\frac{1}{2}\big(\delta^\sigma_\eta
    g^{\mu\nu}W^\rho_{[\mu\nu]}-\delta^\rho_\eta
    g^{\mu\nu}W^\sigma_{[\mu\nu]}\big)\bigg)\\
    &+\bigg[c_1\bigg(g^{\mu\rho}g^{\lambda\sigma}g_{\alpha\eta}+g^{\rho\mu}g^{\sigma\lambda}g_{\eta\alpha}\bigg)+2c_2g^{\mu\lambda}g^{[\rho\sigma]}g_{(\alpha\eta)}\\
    &+\frac{c_3}{2}\bigg(g^{\mu\rho}g^{\sigma\lambda}g_{\alpha\eta}+g^{\rho\mu}g^{\lambda\sigma}g_{\eta\alpha}-g^{\mu\sigma}g^{\rho\lambda}g_{\alpha\eta}-g^{\sigma\mu}g^{\lambda\rho}g_{\eta\alpha}\bigg)\bigg]W^\alpha_{[\mu\lambda]}\\
    &+d\big(g^{(\mu\rho)}\delta^\sigma_\eta-g^{(\mu\sigma)}\delta^\rho_\eta\big)W_\mu\bigg\}=0.
\end{split}
\end{equation}
We contract this equation on $\rho$ and $\eta$ to obtain
\begin{equation} \label{rhosigma}
\begin{split}
    &\bigg(2+8a_1-\frac{3a_2}{2}+b_1\bigg)\partial_\rho(\sqrt{-g}g^{[\sigma\rho]})+\sqrt{-g}\bigg\{\big(b_2+2c_1-c_3-3d\big)g^{(\mu\sigma)}W_\mu\\
    &+b_3g^{[\rho\sigma]}W_\rho-\frac{3}{2}b_3g^{\mu\nu}W^\sigma_{[\mu\nu]}
    +2c_2g^{\mu\nu}g^{[\rho\sigma]}g_{(\alpha\rho)}W^\alpha_{[\mu\nu]}\\
    &-\frac{c_3}{2}\big(g_{\alpha\rho}g^{\mu\sigma}g^{\rho\lambda}+g_{\rho\alpha}g^{\sigma\mu}g^{\lambda\rho}\big)W^\alpha_{[\mu\lambda]}\bigg\}=0
\end{split}
\end{equation}
and the contraction on $\sigma$ and $\eta$ gives
\begin{equation} \label{sigmaeta}
\begin{split}
&3\partial_\eta(\sqrt{-g}g^{(\rho\eta)})
  +\Big(3+2a_1+\frac{3}{2}a_2-b_1\Big)\partial_\eta
g^{[\rho\eta]}
  +\sqrt{-g}\bigg\{3g^{\mu\nu}W^\rho_{\mu\nu}+2g^{\rho\eta}W_\eta\\
&+\Big(3b_1+\frac{3}{2}b_3\Big)g^{[\mu\nu]}W^\rho_{\mu\nu}
  +(2b_1+b_3)g^{[\rho\eta]}W_\eta
  +\Big(-b_2+2c_1+c_3+3d\Big)g^{(\nu\rho)}W_\nu\\
&2c_2g^{[\mu\lambda]}g^{[\rho\eta]}g_{(\alpha\eta)}
 W^\alpha_{\mu\lambda}
 +\frac{c_3}{2}
     \Big(g^{\mu\rho}g^{\eta\lambda}g_{\alpha\eta}
        +g^{\rho\mu}g^{\lambda\eta}g_{\eta\alpha}
      \Big)W^\alpha_{[\mu\lambda]}\bigg\}.
\end{split}
\end{equation}
 We only need to calculate
the connection up to first order, since one can easily check that
all second order contributions of the connection to the lagrangian
(\ref{lag}) can be written as a total derivative. From
(\ref{rhosigma}) we find up to first order that
\begin{equation} \label{W}
W_\mu=\Sigma\nabla^\rho B_{\mu\rho}
\end{equation}
where we have defined
\begin{equation} \label{L}
\Sigma\equiv-\frac{2+8a_1-\frac{3}{2}a_2+b_1}{b_2-2(c1+c3)-3d}.
\end{equation}
In principle one could think that we have special cases if either
the nominator or the denominator of (\ref{L}) is zero.
This means that (on shell) either $W_\mu=0$, or
$\nabla^\rho B_{\mu\rho}=0$. It turns out however, that in the
linearized equations $W_\mu$ and $\nabla^\rho B_{\mu\rho}$
play exactly the same r$\hat{\text{o}}$le. Therefore nothing
special (in comparison to the case where $\Sigma\neq 0$) happens
in the cases where one of them is zero. A case that \emph{is}
special is when both the nominator and the
denominator of (\ref{L}) are zero. However in this case we cannot
proceed with the calculation of the connection. We therefore
assume that this is not the case. We redefine the connection as
follows
\begin{equation} \label{gammadef}
W^\alpha_{\mu\nu}=\Gamma^\alpha_{\mu\nu}-\frac{2}{3}\delta^\alpha_\mu
W_\nu,
\end{equation}
which means that $\Gamma^\alpha_{[\mu\alpha]}=0$, and use
(\ref{rhosigma})) and (\ref{sigmaeta}) to rewrite
(\ref{fullcompat}). Terms that are clearly 2nd order or higher
have been dropped.
\begin{equation}\label{bla}
\begin{split}
&-\partial_\eta
g^{\rho\sigma}+\frac{1}{2}g^{\rho\sigma}g_{\alpha\beta}\partial_\eta
g^{\alpha\beta}-b_1\partial_\eta
g^{[\rho\sigma]}+\frac{b_1}{2}g^{[\rho\sigma]}g_{\alpha\beta}\partial_\eta
g^{\alpha\beta}+g^{\rho\sigma}\Gamma^\lambda_{\eta\lambda}-g^{\mu\sigma}\Gamma^\rho_{\mu\eta}-g^{\rho\mu}\Gamma^\sigma_{\eta\mu}\\
&+b_1\Big(g^{[\rho\sigma]}\Gamma^\lambda_{\eta\lambda}-g^{[\mu\sigma]}\Gamma^\rho_{\mu\eta}-g^{[\rho\mu]}\Gamma^\sigma_{\eta\mu}\Big)\\
&+b_2\Big(g^{(\nu\sigma)}\Gamma^rho_{[\nu\eta]}+g^{(\rho\nu)}\Gamma^\sigma_{[\eta\nu]}\Big)+2(c1+c3)g^{(\mu\rho)}g^{(\nu\sigma)}g_{(\alpha\eta)}\Gamma^\alpha_{[\mu\nu]}\\
&+\mathbb{L}g^{(\rho\nu)}W_\nu\delta^\sigma_\eta+\mathbb{K}\Big(\delta^\sigma_\eta
g^{(\rho\nu)}-\delta^\rho_\eta g^{(\sigma\nu)}\Big)W_\nu
+\mathcal{O}(B^2)=0,
\end{split}
\end{equation}
where we have defined
\begin{equation}
\begin{split}
&\mathbb{L}\equiv\Big(\frac{2}{3}+\frac{4}{3}a_1\Big)
                  \Big(b_2-2(c1+c3)-3d\Big)
                   \Big(2+8a_1-\frac{3}{2}a_2+b_1\Big)^{-1}\\
&\mathbb{K}\equiv \Big(2a_1-\frac{1}{2}a_2\Big)\Big(b_2-2(c1+c3)-3d\Big)
                  \Big(2+8a_1-\frac{3}{2}a_2+b1\Big)^{-1}
                  -\frac{1}{3}b_2+\frac{2}{3}(c_1+c_3)+d.
\end{split}
\nonumber
\end{equation}
\vskip -0.5cm\noindent
Next we multiply (\ref{bla}) with $g_{\rho\sigma}$ and substitute
the resulting expression back in. For convenience we also multiply
our equation with $g_{\rho\beta}g_{\alpha\sigma}$ and obtain:
\begin{equation}
\begin{split}
\partial_\eta
&g_{\alpha\beta}-g_{\rho\beta}\Gamma^\rho_{\alpha\eta}-g_{\alpha\sigma}
\Gamma^\sigma_{\eta\beta}+b_1\nabla_\eta g_{[\alpha\beta]}\\
&-b_2\Big(g_{(\rho\beta)}\Gamma^\rho_{\eta\alpha}+g_{(\alpha\sigma)}\Gamma^\sigma_{[\beta\eta]}\Big)+2(c_1+c_3)g_{(\lambda\eta)}\Gamma^\lambda_{[\beta\alpha]}\\
&+\mathbb{L}W_\beta
g_{\alpha\eta}-\frac{1}{2}\mathbb{L}g_{\alpha\beta}W_\eta+\mathbb{K}\Big(g_{(\alpha\eta)}\delta^\nu_\beta-g_{(\eta\beta)}\delta^\nu_\alpha\Big)W_\nu+\mathcal{O}(B^2)=0.
\end{split}
\end{equation}
The zeroth order solution of this equation clearly gives the well
known  Levi-Civit\`a connection:
\begin{equation}
{}^{(0)}\Gamma^\alpha_{\mu\nu}=\big\{^\alpha_{\mu\nu}\}.
\end{equation}
The first order equation can also be  solved and gives
\begin{equation} \label{connection}
\begin{split}
&{}^{(1)}\Gamma^\rho_{(\eta\beta)}=\frac{\mathbb{L}}{4}\Big(3G^{\lambda\rho}G_{\eta\beta}W_\lambda-\delta^\rho_\beta
W_\eta-\delta^\rho_\eta W_\beta\Big)\\
&{}^{(1)}\Gamma^\rho_{[\eta\beta]}=G^{\rho\zeta}\Big(\phi\nabla_\beta
B_{\eta\zeta}+\psi\nabla_\eta B_{\zeta\beta}+\xi\nabla_\zeta
B_{\beta\eta}\Big)+\theta\bigg(\delta^\rho_\beta
W_\eta-\delta^\rho_\eta W_\beta\bigg)
\end{split}
\end{equation}
where the coefficients in the last expression are given by
\begin{equation}
\begin{split}
&\phi=\psi\equiv\frac{(\mathcal{A}+\mathcal{B})(b_1+1)}{\mathcal{A}^2+\mathcal{A}\mathcal{B}-2\mathcal{B}^2}\\
&\xi\equiv\frac{(\mathcal{A}+3\mathcal{B})(b_1+1)}{\mathcal{A}^2+\mathcal{A}\mathcal{B}-2\mathcal{B}^2}\\
&\theta\equiv\frac{2\mathbb{K}+\mathbb{L}}{\mathcal{A}-\mathcal{B}}\\
&\mathcal{A}\equiv 2(1-b_2+c_1+c_3)\\
&\mathcal{B}\equiv-2(c_1+c_3)
\end{split}
\end{equation}
\subsection{Calculating the linearized lagrangian}
Now that we now the connection, we can calculate the linearized
version of the lagrangian (\ref{lag}). It is clear that the zeroth
order lagrangian is
\begin{equation}
{}^{(0)}\mathcal{L}=\sqrt{-G}(R+2\Lambda)
.
\end{equation}
It is also not difficult to see that (\ref{W}) implies that
\begin{equation}
{}^{(1)}\mathcal{L}=0.
\end{equation}
for the quadratic order we find (all $\Gamma's$ refer from now on
to the first order connection)
\begin{equation}
\begin{split}
{}^{(2)}\mathcal{L}=\sqrt{-G}
 \bigg\{&\bigg(\frac{1}{2}
  \Big(\frac{1}{2}-\rho+2\sigma\Big)B^2G^{\mu\nu}
  +(1-\rho)B^{\mu\alpha}B_\alpha{}^\nu\bigg)R_{\mu\nu}+\frac{1}{4}m^2B^2
\\
&+G^{\mu\nu}\bigg(
              -\Gamma^\sigma_{(\mu\lambda)}\Gamma^\lambda_{(\sigma\nu)}
              +\Gamma^\sigma_{(\mu\nu)}\Gamma^\lambda_{(\sigma\lambda)}
             +(b_2-1)\Gamma^\sigma_{[\mu\lambda]}\Gamma^\lambda_{[\sigma\nu]}\\
&\hskip 1.2cm
+(c_1+c_3)G^{\lambda\delta}G_{\alpha\beta}\Gamma^\alpha_{[\mu\lambda]}
 \Gamma^\beta_{[\nu\delta]}
   +\Big(d-\frac{1}{3}b_2+\frac{2}{3}(c_1+c_3)\Big)W_\mu
W_\nu\bigg)\\
&+B^{\mu\nu}\bigg((1+b_1)\nabla_\lambda\Gamma^\lambda_{[\mu\nu]}
  -\nabla_\nu\Gamma^\lambda_{(\mu\lambda)}
  +\Big(-\frac{2}{3}-\frac{8}{3}a_1+\frac{1}{2}a_2-\frac{1}{3}b_1\Big)
  (\partial_\mu W_\nu-\partial_\nu W_\mu)
\\
&\hskip 1.2cm
  +a_1(\nabla_\mu\Gamma^\lambda_{(\nu\lambda)}
  -\nabla_\nu\Gamma^\lambda_{(\mu\lambda)})\bigg)\bigg\}.
\end{split}
\end{equation}
where the mass for the $B$-field is given by
\begin{equation} \label{mass}
\frac{1}{4}m^2=\Lambda\Big(\frac{1}{2}-\rho+4\sigma\Big)
\end{equation}
 Plugging in the expressions
we found for the connection (\ref{connection}) and removing a
total derivative term results in
\begin{equation}
\begin{split}
{}^{(2)}\mathcal{L}=\sqrt{-G}\bigg\{&\bigg(\frac{1}{2}\big(\frac{1}{2}-\rho+2\sigma\big)B^2G^{\mu\nu}+(1-\rho)B^{\mu\alpha}B_\alpha{}^\nu\bigg)R_{\mu\nu}+\frac{1}{4}m^2B^2\\
&+(\Phi+\frac{1}{3}\Omega)H^2+\Xi(\nabla^\rho
B_{\mu\rho})(\nabla^\lambda
B_{\nu\lambda})G^{\mu\nu}+(\Psi-2\Omega)(\nabla^\sigma
B^{\rho\nu})(\nabla_\rho B_{\nu\sigma})\bigg\}
\end{split}
\end{equation}
where
\begin{equation} \label{lag2}
\begin{split}
&H_{\alpha\beta\gamma}=\partial_\alpha
B_{\beta\gamma}+\partial_\beta B_{\gamma\alpha}+\partial_\gamma
B_{\alpha\beta}\\
&\Phi=\frac{1}{3}(\phi^2+2\xi\phi)(b_2+c_1+c_3-1)\\
&\Xi=3\Sigma^2\theta^2\big(1-b_2+2(c1+c3)\big)
   +\Sigma^2\Big(d-\frac{1}{3}b_2+\frac{2}{3}(c_1+c_3)
                -\frac{3}{8}\mathbb{L}^2
            \Big)
\\
&\quad\quad
  +\Sigma\bigg(2\Big(\theta+b_1\theta-\frac{2}{3}
                    -\frac{8}{3}a_1+\frac{1}{2}a_2-\frac{1}{3}b_1
                    -\frac{\mathbb{L}}{2}(a_1+\frac{1}{2})
                \Big)-\theta(\phi-\xi)(4(c_1+c_3)-2b_2+2)\bigg)\\
&\Psi=(b_2-1)(\xi-\phi)^2+2\phi(1+b_1)\\
&\Omega=(c_1+c_3)(\xi-\phi)^2+\xi(1+b_1).
\end{split}
\end{equation}
 van Nieuwenhuizen has proven a theorem~\cite{Nieuwenhuizen:1973},
which states that in flat space the
only consistent action for a massive antisymmetric tensor field is
of the form
\begin{equation} \label{nieuwenhuizen}
{\cal S}_{\rm B\, flat} = \int d^4 x \Big(-\frac{1}{12}H^2+\frac{1}{4}m^2B^2\Big)
\,.
\end{equation}
If we want to make sure that (\ref{lag2}) reduces to this form
in flat space, the terms with the covariant derivatives have to be
combined to a curvature tensor. This can be done with the
following identity
\begin{equation}
(\nabla^\mu B^{\nu\beta})(\nabla_\beta B_{\mu\nu})+(\nabla_\rho
B^{\mu\rho})(\nabla^\sigma
B_{\mu\sigma})=-B^{\mu\alpha}B_\alpha{}^\nu
R_{\mu\nu}-B^{\mu\nu}B^{\alpha\beta}R_{\mu\alpha\nu\beta}+\text{total
derivative}
\end{equation}
With this identity we find that the requirement
(\ref{nieuwenhuizen}) is satisfied if
\begin{equation} \label{nieuwconstr}
\begin{split}
    &\Phi+\frac{1}{3}\Omega=-\frac{1}{12}\\
    &\Xi=\Psi-2\Omega
\end{split}
\end{equation}
and then the lagrangian reduces to its final form
\begin{equation}
\begin{split}
    \mathcal{L}=\sqrt{-G}
           \Big[&R+2\Lambda-\frac{1}{12}H^2
               +\Big(\frac{1}{4}m^2+\beta R\Big)B^2
\\
           &-\alpha R_{\mu\nu}B^{\mu\alpha}B_\alpha{}^\nu-\gamma
              R_{\mu\alpha\nu\beta}B^{\mu\nu}B^{\alpha\beta}
         \Big]+\mathcal{O}(B^3),
\end{split}
\end{equation}
where
\begin{equation}
\begin{split}
    &\alpha=\rho+\Xi-1\\
    &\beta=\frac{1}{2}\Big(\frac{1}{2}-\rho+2\sigma\Big)\\
    &\gamma=\Xi
\end{split}
\end{equation}
One can show that there is no choice for the
parameters in the full lagrangian that has $\Xi=0$ and
simultaneously satisfies the constraints (\ref{nieuwconstr}) (what
one finds is that if $\Xi=\Psi-2\Omega=0$, then automatically
$\Phi+\frac{1}{3}\Omega=0$). Therefore we conclude that it is
not possible to get rid of the $B$-field coupling to the Riemann tensor.
In other words, $\gamma=0$ is not allowed within the theory.

\section{Geometric quantities} \label{ageom}
We give an overview of the geometric quantities we use in the
text. For the FRLW case we have
\begin{equation}
\begin{split}
    &\big\{{}^0_{00}\big\}=\mathcal{H},\quad\quad\big\{{}^0_{ij}\big\}=\mathcal{H}\delta_{ij},\quad\quad
    \big\{{}^j_{0i}\big\}=\mathcal{H}\delta^j_i\\
    &R_{00}=-3\mathcal{H}',\quad\quad\quad\quad\quad
    R^0_0=-\frac{3}{a^2}\mathcal{H}'\\
    &R_{ij}=(\mathcal{H}'+2\mathcal{H}^2)\delta_{ij},\quad\quad
    R^j_i=-\frac{1}{a^2}(\mathcal{H}'+2\mathcal{H}^2)\delta^j_i\\
    &R=-\frac{6}{a^2}(\mathcal{H}^2+\mathcal{H}').
\end{split}
\end{equation}
For the Schwarzschild case we have (the index $a$ denotes 2,3):
\begin{equation}
\begin{split}
    &R^0{}_\mu{}^1{}_\nu=\delta^1_\mu\delta^0_\nu\frac{2M_s}{r^3}\\
    &R^0{}_\mu{}^a{}_\nu=-\delta^a_\mu\delta^0_\nu\frac{M_s}{r^3}\\
    &R^1{}_\mu{}^a{}_\nu=-\delta^a_\mu\delta^1_\nu\frac{M_s}{r^3}\\
    &R^2{}_\mu{}^3{}_\nu=\delta^3_\mu\delta^2_\nu\frac{2M_s}{r^3}\\
\end{split}
\end{equation}
The terms of the form $R^\lambda{}_\mu{}^\lambda{}_\nu$ (no
summation over $\lambda$) are symmetric in $\mu$ and $\nu$ and
therefore they do not contribute to our equations.

\bibliographystyle{utcaps}
\bibliography{biblio}

\end{document}